\documentclass[a4paper,12pt]{article}
\usepackage{amsfonts,amssymb,graphicx}

\font\litera=pplr at 12pt
\font\literaTitlu=pplr scaled \magstep4
\font\literaTitluSect=pplr scaled \magstep3
\def\pal{\let\rm=\litera\baselineskip=16pt\rm}
\def\palTitlu{\let\rm=\literaTitlu\rm}
\def\palTitluSect{\let\rm=\literaTitluSect\rm}

\begin{document}

\author{
Octavian G. Mustafa\\
\small{University of Craiova, DAL,}\\
\small{Department of Mathematics {\&} Computer Science,}\\
\small{A.I. Cuza 13, Craiova, Romania}\\
\small{e-mail address: octawian@yahoo.com}
}

\title{On a recent model of tsunami background flow}
\date{}
\maketitle

\noindent{\bf Abstract} We present a class of vorticity functions that will allow for isolated, circular vorticity regions in the background of still water preceding the arrival of a tsunami wave at the shoreline.

\section{Introduction}
In two seminal papers \cite{cj1,cj2}, Constantin and Johnson proposed a model for studying what happens beneath the surface of the ocean before the arrival of a destructive tsunami wave at the shoreline. As opposed to other enterprises, the fluid does not move here irotationally beneath its surface in a global manner, but the water is still with the exception of some isolated, bounded regions where it moves with vorticity. Given the particular character of the vorticity region discussed in \cite{cj1}, the authors envisaged the possibility of more permissive shapes for the boundary of such regions in \cite{cj2}: the circular vorticity region. The analysis has been put on a firm ground via a dynamical systems approach in the paper \cite{constantin1}.

Our intention in this note is to investigate the key features of the technical proof from the latter work and, by relaxing them, to propose a new candidate for the vorticity function in the circular vorticity region. In a loose manner, {\it the new function is a slight deviation from the Constantin-Johnson vorticity\/}.
\section{Phase-plane analysis}
Following \cite[Eq. (3.1)]{constantin1}, let us introduce the nonlinear ODE system
\begin{eqnarray}
\left\{
\begin{array}{ll}
\psi^{\prime}=\beta,\\
\beta^{\prime}=-\frac{\beta}{r}-f(\psi),
\end{array}
\right.\quad r>0,\label{main_ODE}
\end{eqnarray}
where $f:\mathbb{R}\rightarrow\mathbb{R}$ is a continuous (vorticity) function with some extra restrictions to be described in the following. First of all, $f$ has {\it three zeros\/}: $0$, $u_{0}\in(0,1]$ and $u_{1}=-u_{0}$. See Figure 1.

{\bf The energy and its decay.\/} We define the {\lq\lq}energy{\rq\rq} $E(\psi,\beta)=\frac{\beta^{2}}{2}+\int_{0}^{\psi}f(u)du$ and ask also that $f$ be {\it odd\/}, meaning that $E(\pm\psi,\pm\beta)=E(\psi,\beta)$, together with $\lim\limits_{u\rightarrow\pm\infty}f(u)=\pm\infty$. The latter restriction leads to 
\begin{eqnarray}
\lim\limits_{\psi\rightarrow\pm\infty}\frac{1}{\psi}\int_{0}^{\psi}f(u)du=+\infty\label{marg_energ}
\end{eqnarray}
by means of the L'H\^{o}pital rule.

\begin{figure*}[h]
    \centering
        \includegraphics[width=11cm,height=7cm]{C:/octavian/poza1_viena.bmp}\\
        {\pal Figure 1}
    \label{fig:unu}
\end{figure*}

If $E(r)=E(\psi(r),\beta(r))$ for some solution $(\psi,\beta)$ of the system (\ref{main_ODE}), with $r\in(0,r_{\infty})$ and $r_{\infty}\leq+\infty$, then
\begin{eqnarray*}
E^{\prime}(r)=\beta\beta^{\prime}+f(\psi)\psi^{\prime}=-\frac{\beta^{2}}{r},\quad r>0.
\end{eqnarray*}
This means that the energy $E$ is monotone non-increasing along the solution $(\psi,\beta)$. Obviously, for the three equilibria  $(0,0)$, $(u_{0},0)$, $(u_{1},0)$, we have constant energy throughout $(0,+\infty)$: $0$ for the null solution, and $\int_{0}^{u_{0}}f(u)du=\int_{0}^{u_{1}}f(u)du$ for the other two. To ensure that {\it the energy $E(r)$ is strictly decreasing along any non-equilibrium solution $(\psi,\beta)$ whenever it is positive\/}, one way is to impose that $\int_{0}^{u_{i}}f(u)du<0$, where $i\in\{0,1\}$. This (strict) monotonicity restriction is a key feature of the estimate \cite[Eq. (3.13)]{constantin1}. In our case, following Constantin's technique, if $E(r_{1})=E(r_{2})>0$, which reads as $0=E(r_{2})-E(r_{1})=-\int_{r_{1}}^{r_{2}}\frac{\beta^{2}}{r}dr$ for some $r_{1}<r_{2}$ from $(0,r_{\infty})$, then $\beta\equiv0$ in $[r_{1},r_{2}]$. Taking into account the second of equations (\ref{main_ODE}), we get $f(\psi(r))=0$ throughout $[r_{1},r_{2}]$ which means that either $\psi=u_{0}$ or $\psi=u_{1}$ in $[r_{1},r_{2}]$. However, for any of these solutions  we have $E(r_{1})=E(r_{2})<0$, which is a contradiction.

Another way of establishing this decay of the energy along non-trivial solutions is to ask that {\it$f$ be $C^{1}$ in $\mathbb{R}-\{0\}$\/}, as in the case of Constantin's vorticity function, namely
\begin{eqnarray}
f(u)=\left\{
\begin{array}{ll}
u-\frac{u}{\sqrt{\vert u\vert}},\quad u\neq0,\\
0,\quad u=0.
\end{array}
\right.\label{vorticitatea_constantin}
\end{eqnarray}
Now, given $(\psi,\beta)$ a non-trivial and non-equilibrium solution, that is $\psi^{2}+\beta^{2}>0$ everywhere in $(0,r_{\infty})$ and $\psi(r)\not\in\{0,u_{0},u_{1}\}$ when $\beta(r)=0$, we have
\begin{eqnarray}
E^{\prime\prime}=-\frac{2\beta\beta^{\prime}}{r}+\frac{\beta^2}{r^2}=\frac{\beta}{r}\left[\frac{3\beta}{r}+f(\psi)\right],\label{eprim}
\end{eqnarray}
and
\begin{eqnarray}
E^{\prime\prime\prime}=-\frac{2}{r}\cdot(\beta^{\prime})^{2}+\frac{2\beta}{r}\left(-\beta^{\prime\prime}+\frac{2\beta^{\prime}}{r}-\frac{\beta}{r^2}\right).\label{eprimm}
\end{eqnarray}
Obviously, $\beta^{\prime\prime}$ exists for all the values $r>0$ with $\psi(r)\neq0$. Since the solution $(\psi,\beta)$ is continuous, then in the (small) vicinity of any $r_{+}>0$ such that $\beta(r_{+})=0$, meaning $\psi(r_{+})\not\in\{0,u_{0},u_{1}\}$, the function $\psi$ has constant, non-null, sign. So, by means of (\ref{eprim}), (\ref{eprimm}), we get $E^{\prime}(r_{+})=E^{\prime\prime}(r_{+})=0$, $E^{\prime\prime\prime}(r_{+})<0$. The Taylor expansion $E(r_{+}+h)=E(r_{+})+E^{\prime}(r_{+})h+E^{\prime\prime}(r_{+})\frac{h^2}{2}+E^{\prime\prime\prime}(r_{+})\frac{h^3}{6}+o(\vert h\vert^{3})$ yields $E(r_{+}+h)<E(r_{+})$ when $h^{3}\equiv0$ and $h>0$. If $\beta(r)\neq0$ then, obviously, the energy $E$ decreases strictly on a small right-neighborhood of $r$.

{\bf Local existence I.\/} Assume that the continuous function $f$ from (\ref{main_ODE}) satisfies the following restriction: for any $a\geq1$ there exists $\eta=\eta(a)\in\left(3,\frac{7}{2}\right]$ such that
\begin{eqnarray}
\vert f(\xi)\vert\leq\eta a\quad\mbox{when }\xi\in\left[\left(1-\frac{\eta}{4}\right)a,\left(1+\frac{\eta}{4}\right)a\right].\label{estimare_aux_01}
\end{eqnarray}
In particular, for the vorticity function from (\ref{vorticitatea_constantin}) we have $\vert f(\xi)\vert\leq\vert\xi\vert+\sqrt{\vert\xi\vert}\leq\left(1+\frac{\eta}{4}+\sqrt{1+\frac{\eta}{4}}\right)a\leq\eta a$ whenever $a\geq1$ and $\eta\geq\frac{28}{9}$.

We claim that the integral equation
\begin{eqnarray}
\psi(r)=a-\int_{0}^{r}\frac{1}{\xi}\int_{0}^{\xi}\tau f(\psi(\tau))d\tau d\xi,\quad r\geq0,\label{ec_integrala_1}
\end{eqnarray}
has a solution in $X=C([0,1],\mathbb{R})$. The proof of our claim is based on the Schauder fixed point theorem and on the Ascoli-Arzela relative compactness criterion. Notice first that, by means of the L'H\^{o}pital rule, we have $\lim\limits_{\xi\searrow0}\frac{1}{\xi}\int_{0}^{\xi}\tau f(\psi(\tau))d\tau=\lim\limits_{\xi\searrow0}[\xi f(\psi(\xi))]=0$, which means that the function $\xi\mapsto\frac{1}{\xi}\int_{0}^{\xi}\tau f(\psi(\tau))$ can be prolongated backwards to zero as a continuous function.

Now, consider $D$ the closed ball of radius $\frac{\eta a}{4}$ and center $a$  of the Banach space ${\cal B}=(X,\Vert\cdot\Vert_{\infty})$. Obviously, the ball is convex. Let $T:D\rightarrow X$ be an operator with $T(\psi)$ given by the right-hand member of the equation (\ref{ec_integrala_1}) for any $\psi\in D$. We have the estimates: {\it(i) boundedness of $T(D)$\/}, namely
\begin{eqnarray*}
\vert T(\psi)(r)-a\vert\leq\int_{0}^{r}\frac{1}{\xi}\int_{0}^{\xi}\tau\cdot\eta a\,d\tau d\xi=\frac{\eta a}{4}\cdot r\leq\frac{\eta a}{4},\quad\psi\in D.
\end{eqnarray*}
Remark that we have obtained actually that $T(D)\subseteq D$. {\it(ii) Equicontinuity of $T(D)$\/}, which follows from
\begin{eqnarray*}
\vert [T(\psi)]^{\prime}(r)\vert\leq\frac{1}{r}\int_{0}^{r}\tau\cdot\eta a\,d\tau\leq\frac{\eta a}{2},\quad r\in(0,1].
\end{eqnarray*}
According to the Ascoli-Arzela criterion, the set $T(D)$ is relatively compact in ${\cal B}$, yielding that the operator $T$ transports bounded sets into relatively compact ones. 

Finally, {\it(iii) the operator $T$ is continuous\/}. Since the function
\begin{eqnarray*}
f:\left[\left(1-\frac{\eta}{4}\right)a,\left(1+\frac{\eta}{4}\right)a\right]\rightarrow\mathbb{R}
\end{eqnarray*}
is uniformly continuous, for every $\varepsilon>0$ there exists $\delta=\delta(\varepsilon)>0$ such that, given $\psi_{1}$, $\psi_{2}\in D$ with $\Vert\psi_{1}-\psi_{2}\Vert_{\infty}\leq\delta$, one has $\vert f(\psi_{1}(r))-f(\psi_{2}(r))\vert\leq\varepsilon$, $r\in[0,1]$. We have
\begin{eqnarray*}
\vert T(\psi_{1})(r)-T(\psi_{2})(r)\vert\leq\int_{0}^{r}\frac{1}{\xi}\int_{0}^{\xi}\tau\cdot\varepsilon\,d\tau d\xi\leq\frac{\varepsilon}{4},\quad r\in[0,1].
\end{eqnarray*}

According to the Schauder fixed point theorem, the operator $T$ has a fixed point $\psi\in D$. In particular, $\psi(0)=a$. Also, $\psi\in C^{1}((0,1],\mathbb{R})$ and $\psi^{\prime}(r)=\frac{1}{r}\int_{0}^{r}\tau f(\psi(\tau))d\tau$, $r>0$. By means of L'H\^{o}pital rule, $\lim\limits_{r\searrow0}\psi^{\prime}(r)=0$. This means that $\psi\in C^{1}([0,1],\mathbb{R})$ verifies (\ref{ec_integrala_1}) and, given its smoothness, yields the solution $(\psi,\psi^{\prime})\in C([0,1],\mathbb{R}^{2})$ of the system (\ref{main_ODE}) with the starting point $(\psi(0),\psi^{\prime}(0))=(a,0)$. In particular, by being an element of $D$, the function $\psi$ satisfies the inequality
\begin{eqnarray}
\psi(r)\geq\left(1-\frac{\eta(a)}{4}\right)a\geq\frac{a}{8}.\label{estimare_aux_00}
\end{eqnarray}

{\bf Local existence II.\/} We are interested here in the {\it backward\/} existence and uniqueness of the solution to (\ref{main_ODE}). Assume that, in addition to (\ref{estimare_aux_01}), there exists $L=L(a)\in\left(0,\frac{5}{2}\right)$ such that $\vert f(\xi_{1})-f(\xi_{2})\vert\leq L\vert\xi_{1}-\xi_{2}\vert$ for all $\xi_1$, $\xi_{2}\in\left[\left(1-\frac{\eta}{4}\right)a,\left(1+\frac{\eta}{4}\right)a\right]$. In the particular case of (\ref{vorticitatea_constantin}), we have $\vert f^{\prime}(\xi)\vert\leq 1+\frac{1}{2\sqrt{\xi}}\leq 1+\frac{1}{\sqrt{4\left(1-\frac{\eta}{4}\right)a}}\leq 1+\sqrt{\frac{2}{a}}\leq 1+\sqrt{2}<\frac{5}{2}$.

Take $T\geq6$ and introduce the system of integral equations
\begin{eqnarray}
\left\{
\begin{array}{ll}
\psi(r)=\psi_{T}-\int_{r}^{T}\beta(s)ds,\\
\beta(r)=\frac{\beta_{T}T}{r}+\frac{1}{r}\int_{r}^{T}sf(\psi(s))ds,
\end{array}
\right.\quad r\leq T,\quad\psi_{T},\thinspace\beta_{T}\in\mathbb{R}.\label{sistem_auxiliar_00}
\end{eqnarray}
Here, $\psi_{T}=a$, $\vert\beta_{T}\vert\leq\frac{\eta a}{8}$. For $X=C([\sqrt{T^{2}-1},T],\mathbb{R}^{2})$, let $D=\{(\psi,\beta)\in X:\vert\psi(r)-a\vert\leq\frac{\eta a}{4},\thinspace \vert \beta(r)\vert\leq2\vert\beta_{T}\vert+\eta a,\thinspace r\in[\sqrt{T^{2}-1},T]\}$. Obviously, $D$ is a closed, convex, with non-void interior subset of ${\cal B}=(X,\Vert\cdot\Vert_{\infty})$, where $\Vert(\psi,\beta)\Vert_{\infty}=\max\{\Vert \psi\Vert_{\infty},\Vert \beta\Vert_{\infty}\}$.

Set $\lambda\in(1,3)$ and $k>0$ such that
\begin{eqnarray}
&&(T+\sqrt{T^{2}-1})\ln\lambda\qquad(>11\cdot\ln\lambda)\nonumber\\
&&=\frac{\ln\lambda}{T-\sqrt{T^{2}-1}}>k>\max\left\{\lambda\ln\lambda,L\left(\frac{5}{4}\lambda\ln\lambda+\frac{1}{2}\right)\right\}\label{estimare_aux_02}
\end{eqnarray}
and introduce the metric
\begin{eqnarray*}
&&d((\psi_{1},\beta_{1}),(\psi_{2},\beta_{2}))\\
&&=\sup\limits_{r\in[\sqrt{T^{2}-1},T]}\left[\max\{\vert\psi_{1}(r)-\psi_{2}(r)\vert,\vert\beta_{1}(r)-\beta_{2}(r)\vert\}\cdot\mbox{e}^{-kr}\right].
\end{eqnarray*}
Notice the by-product inequality $11\cdot\ln\lambda>L\left(\frac{5}{4}\cdot3\ln\lambda+\frac{1}{2}\right)$. This is equivalent to $\frac{44\ln\lambda}{15\ln\lambda+2}>L$. Since the function $x\mapsto\frac{44x}{15x+2}$ is increasing in $(0,\ln3)$ and $\frac{44\ln3}{15\ln3+2}>\frac{5}{2}>L(a)$, such a constant $\lambda$ exists always. 

Then, ${\cal M}=(D,d)$ constitutes a complete metric space. Introduce further the operator $T:D\rightarrow X$ with $T(\psi,\beta)$ given by the right-hand member of the system (\ref{sistem_auxiliar_00}) for any $(\psi,\beta)\in D$: $T(\psi,\beta)=(T_{1}(\psi,\beta),T_{2}(\psi,\beta))$. 

We have the estimates
\begin{eqnarray*}
\vert T_{2}(\psi,\beta)(r)\vert&\leq&\vert\beta_{T}\vert\cdot\frac{T}{\sqrt{T^{2}-1}}+\frac{T}{\sqrt{T^{2}-1}}\int_{r}^{T}\eta a\,ds\\
&\leq&2\vert\beta_{T}\vert+2\eta a(T-\sqrt{T^{2}-1})=2\vert\beta_{T}\vert+\frac{2\eta a}{T+\sqrt{T^{2}-1}}\\
&<&2\vert\beta_{T}\vert+\eta a
\end{eqnarray*}
and respectively
\begin{eqnarray*}
\vert T_{1}(\psi,\beta)(r)-a\vert&\leq&\int_{r}^{T}(2\vert\beta_{T}\vert+\eta a)ds\leq\frac{2\vert\beta_{T}\vert+\eta a}{T+\sqrt{T^{2}-1}}\\
&<&\frac{2\cdot\frac{\eta a}{8}+\eta a}{5}=\frac{\eta a}{4}.
\end{eqnarray*}
They read as $T(D)\subseteq D$.

Given $(\psi_{i},\beta_{i})\in D$, $i\in\{1,2\}$, we get
\begin{eqnarray*}
\vert T_{1}(\psi_{1},\beta_{1})(r)-T_{1}(\psi_{2},\beta_{2})(r)\vert&\leq&\int_{r}^{T}\mbox{e}^{ks}ds\cdot d((\psi_{1},\beta_{1}),(\psi_{2},\beta_{2}))\\
&=&\mbox{e}^{kr}\cdot\frac{\mbox{e}^{k(T-r)}-1}{k}\cdot d((\psi_{1},\beta_{1}),(\psi_{2},\beta_{2})).
\end{eqnarray*}
Recall the elementary inequality $\mbox{e}^{x}\leq 1+\lambda x$ for any $\lambda>1$ and $x\in[0,\ln\lambda]$. We have $x=k(T-r)\leq k(T-\sqrt{T^{2}-1})<\ln\lambda$ by means of the first of inequalities (\ref{estimare_aux_02}). So, $\mbox{e}^{k(T-r)}-1\leq\lambda k(T-r)$ and we get
\begin{eqnarray*}
\vert T_{1}(\psi_{1},\beta_{1})(r)-T_{1}(\psi_{2},\beta_{2})(r)\vert\cdot\mbox{e}^{-kr}&\leq&\lambda(T-r)
\cdot d((\psi_{1},\beta_{1}),(\psi_{2},\beta_{2}))\\
&\leq&\frac{\lambda\ln\lambda}{k}\cdot d((\psi_{1},\beta_{1}),(\psi_{2},\beta_{2})),
\end{eqnarray*}
where $r\in[\sqrt{T^{2}-1},T]$.

Further,
\begin{eqnarray*}
\vert T_{2}(\psi_{1},\beta_{1})(r)-T_{1}(\psi_{2},\beta_{2})(r)\vert&\leq&\frac{L}{r}\int_{r}^{T}\tau\mbox{e}^{k\tau}d\tau\cdot d((\psi_{1},\beta_{1}),(\psi_{2},\beta_{2}))
\end{eqnarray*}
and, since \begin{eqnarray*}
\frac{1}{r}\int_{r}^{T}\tau\mbox{e}^{k\tau}d\tau&=&\frac{T\mbox{e}^{kT}-r\mbox{e}^{kr}}{kr}-\frac{1}{kr}\int_{r}^{T}\mbox{e}^{k\tau}d\tau\leq\frac{T\mbox{e}^{kT}-r\mbox{e}^{kr}}{kr}\\
&\leq&\mbox{e}^{kr}\cdot\frac{T\mbox{e}^{k(T-r)}-\sqrt{T^{2}-1}}{k\sqrt{T^{2}-1}}<\mbox{e}^{kr}\cdot\frac{T\mbox{e}^{k(T-r)}-(T-1)}{k\sqrt{T^{2}-1}}\\
&\leq&\mbox{e}^{kr}\cdot\frac{T\cdot\lambda k(T-r)+1}{k\sqrt{T^{2}-1}}<\mbox{e}^{kr}\cdot\frac{T\cdot\lambda\ln\lambda+1}{k\sqrt{T^{2}-1}}\\
&<&\mbox{e}^{kr}\cdot\frac{\frac{5}{4}\sqrt{T^{2}-1}\cdot\lambda\ln\lambda+\frac{\sqrt{T^{2}-1}}{2}}{k\sqrt{T^{2}-1}}=\mbox{e}^{kr}\cdot\frac{\frac{5}{4}\lambda\ln\lambda+\frac{1}{2}}{k},
\end{eqnarray*}
we conclude that
\begin{eqnarray*}
d(T(\psi_{1},\beta_{1}),T(\psi_{2},\beta_{2}))\leq\zeta\cdot d((\psi_{1},\beta_{1}),(\psi_{2},\beta_{2})),
\end{eqnarray*}
where
\begin{eqnarray*}
\zeta=\frac{\max\left\{\lambda\ln\lambda,L\left(\frac{5\lambda\ln\lambda}{4}+\frac{1}{2}\right)\right\}}{k}<1
\end{eqnarray*}
by means of the second of inequalities (\ref{estimare_aux_02}). Being a contraction, the operator $T:{\cal M}\rightarrow{\cal M}$ has a unique fixed point $(\psi,\beta)\in D$ which is the solution of system (\ref{sistem_auxiliar_00}) in $[\sqrt{T^{2}-1},T]$. See Figure 2.

\begin{figure*}[h]
    \centering
        \includegraphics[width=7cm,height=5cm]{C:/octavian/poza2_viena.bmp}\\
        {\pal Figure 2}
    \label{fig:doi}
\end{figure*}

In the particular case when $\psi_{T}=a=u_{0}=1$ and $\beta_{T}=0$, see e.g., the vorticity function (\ref{vorticitatea_constantin}), the uniqueness of solution leads to $\psi\equiv u_{0}$ in $[\sqrt{T^{2}-1},T]$. This remark has an essential consequence: {\it if there exists a non-equilibrium solution $(\psi,\beta)$ of system (\ref{main_ODE}) that will eventually reach the equilibrium $(u_{0},0)$ then either the equilibrium is attained in $r<6$ {\lq\lq}units of time{\rq\rq} or it is attained in {\lq\lq}infinite time{\rq\rq}, that is $\lim\limits_{r\rightarrow+\infty}(\psi(r),\beta(r))=(u_{0},0)$\/}. In fact, let us assume that $[\psi(r)]^{2}+[\beta(r)]^{2}>u_{0}^{2}$ in $[0,6]$ and there is some $T_{0}>6$ with $\psi(T_{0})=u_{0}$ and $\beta(T_{0})=0$. Introduce $T_{-}=\inf\{T>6:\psi(T)=u_{0},\thinspace\beta(T)=0\}$. Then, $T_{-}\geq6$ and $\psi(T_{-})=u_{0}$, $\beta(T_{-})=0$. Thus, the solution $(\psi,\beta)$ is constant in $[\sqrt{T_{-}^{2}-1},T_{-}]$, and so we have $\psi(\sqrt{T_{-}^{2}-1})=u_{0}$, $\beta(\sqrt{T_{-}^{2}-1})=0$, which contradicts the definition of $T_{-}$.

{\bf Polar coordinates (the Pr\"{u}fer transform) I.\/} Let $(\psi,\beta)$ be a solution of the system (\ref{main_ODE}) starting from $(a,0)$, with $a\geq1$. The solution exists throughout $[0,1]$ and we have the estimate $\psi(1)\geq\frac{a}{8}$ by means of (\ref{estimare_aux_00}). Since $[\psi(1)]^{2}+[\beta(1)]^{2}>0$, we introduce the new variables
\begin{eqnarray*}
\psi=R\cos\theta,\quad\beta=R\sin\theta,\quad R=R(r),\thinspace\theta=\theta(r),
\end{eqnarray*}
which make sense on a small right-neighborhood of $r=1$ via the Peano existence theorem for (\ref{main_ODE}). Since $E(r)\leq E(1)$ for as long as the solution $(\psi,\beta)$ exists to the right of $1$ (recall that $E^{\prime}(r)\leq0$), the estimate (\ref{marg_energ}) shows that both of the quantities $\psi(r)$, $\beta(r)$ are bounded on their maximal interval of existence. According to Wintner's non-local existence theorem, this implies {\it the global existence in the future\/} for the solution $(\psi,\beta)$.

In polar coordinates, the system (\ref{main_ODE}) reads as
\begin{eqnarray}
\left\{
\begin{array}{ll}
\theta^{\prime}=-1-\frac{1}{2r}\cdot\sin2\theta+\left[\cos^{2}\theta-\frac{f(R\cos\theta)\cos\theta}{R}\right],\\
R^{\prime}=-\frac{R}{r}\cdot\sin^{2}\theta+[R\cos\theta-f(R\cos\theta)]\sin\theta,
\end{array}
\right.\quad r\geq1.\label{main_ODE_polar}
\end{eqnarray}

We are interested here in generalizing \cite[Eq. (3.13)]{constantin1}. To this end, assume that there exists an {\it odd\/} continuous function $g:\mathbb{R}\rightarrow\mathbb{R}$ such that $f(\psi)=\psi-g(\psi)$ and $\int_{0}^{\psi}g(u)du\geq\frac{\psi g(\psi)}{2\lambda}$, $\psi\in\mathbb{R}$, for some $\lambda\in\left(\frac{1}{2},1\right)$. In particular, for (\ref{vorticitatea_constantin}), $g(\psi)=\sqrt{\vert\psi\vert}\cdot\mbox{sgn}\,\psi$ and $\lambda=\frac{3}{4}$. Also, for the vorticity function $f(\psi)=\psi-\vert \psi\vert^{\alpha}\cdot\mbox{sgn}\,\psi$, where $\alpha\in(0,1)$, we have $\lambda=\frac{1+\alpha}{2}$. 

The first of equations (\ref{main_ODE_polar}) reads as
\begin{eqnarray}
\theta^{\prime}=-1-\frac{\sin2\theta}{2r}+\frac{\psi g(\psi)}{R^{2}},\quad r\geq1.\label{ec_dif_polar_00}
\end{eqnarray}

Assume that $E(r)=E(\psi(r),\beta(r))>0$ for $r\in[1,r_{++})$, where $r_{++}\leq+\infty$, and $\psi g(\psi)\geq0$ for all $\psi\in\mathbb{R}$. Then, we have the estimates
\begin{eqnarray*}
0<E(r)=\frac{\beta^{2}+\psi^{2}}{2}-\int_{0}^{\psi}g(u)du\leq\frac{R^2}{2}-\frac{\psi g(\psi)}{2\lambda},
\end{eqnarray*}
which, via (\ref{ec_dif_polar_00}), lead to
\begin{eqnarray}
-1-\frac{1}{2r}\leq\theta^{\prime}(r)\leq-(1-\lambda)+\frac{1}{2r},\quad r\in[1,r_{++}).\label{estimare_aux_polar_00}
\end{eqnarray}

To add some geometric flavor to this analysis, notice that the {\lq\lq}problematic{\rq\rq} term in the first of equations (\ref{main_ODE_polar}), that is $\frac{\psi g(\psi)}{R^2}$, reads for (\ref{vorticitatea_constantin}) as $\frac{\vert\cos\theta\vert^{3/2}}{\sqrt{R}}$, see \cite[Eq. (3.11)]{constantin1}. Let $(\iota(r)\psi(r),\iota(r)\beta(r))$ be the point of intersection between the algebraic curve $E(\psi,\beta)=0$ and the straight line passing through the origin $O$ of the phase plane $O\psi\beta$ and the current point of the flow (solution) $(\psi(r),\beta(r))$, $r\in[1,r_{++})$. See Figure 1. Thus, $\iota=\frac{16}{9}\cdot\frac{\psi^3}{R^{4}}$, and so we have $\sqrt{\vert\iota\vert}=\frac{\vert\cos\theta\vert^{3/2}}{\sqrt{R}}$.

{\bf Polar coordinates II.\/} Let us impose a new restriction of the last term of (\ref{ec_dif_polar_00}), that is
\begin{eqnarray}
-\frac{c}{R^{\nu}}\leq\frac{\psi g(\psi)}{R^{2}}\leq\frac{1+c}{R^{\nu}},\quad\vert\psi\vert\leq R,\label{estimare_aux_polar_01}
\end{eqnarray}
for some $c\in[0,1)$, $\nu\in(0,1)$. In the particular case of (\ref{vorticitatea_constantin}), we have $c=0$ and $\nu=\frac{1}{2}$.

Take $\delta>\varepsilon>(1+c)^{1/\nu}-1\geq0$ and $a>8(1+\delta)$. We are interested here in the eventual encounter of the trajectory $\{(\psi(r),\beta(r)):r\geq0\}$ starting from $(a,0)$ with the ring $1+\varepsilon<R<1+\delta$, see Figure 1. According to (\ref{estimare_aux_00}), this encounter can happen only for $r\geq1$, so we can use the equation (\ref{ec_dif_polar_00}) to investigate it. We claim that $\liminf\limits_{r\rightarrow+\infty}R(r)\leq(1+c)^{1/\nu}$, meaning {\it the solution will reach the ring in {\lq\lq}finite time{\rq\rq}.\/}

For the sake of contradiction, suppose that $\liminf\limits_{r\rightarrow+\infty} R(r)>1+\varepsilon$. Then, by means of (\ref{ec_dif_polar_00}), we have
\begin{eqnarray*}
-\frac{3}{2}-\frac{c}{(1+\varepsilon)^{\nu}}&\leq&-1-\frac{1}{2r}-\frac{c}{R^{\nu}}\leq\theta^{\prime}\\
&\leq&-1+\frac{1}{2r_{-}}+\frac{1+c}{(1+\varepsilon)^{\nu}},\quad r\geq r_{-}\geq1.
\end{eqnarray*}
Obviously, $R(r)>1+\varepsilon$ for all $r\geq r_{-}$. Since $(1+c)(1+\varepsilon)^{-\nu}<1$, we can take {\it$r_{-}$ great enough\/} to have $-1+\frac{1}{2r_{-}}+(1+\varepsilon)^{-\nu}=-\eta<0$. By now, $-\frac{3}{2}-\frac{c}{(1+\varepsilon)^{\nu}}\leq\theta^{\prime}\leq-\eta$ throughout $[r_{-},+\infty)$.

Recalling (\ref{marg_energ}), observe that the quantity $E(r)$ is bounded from below and monotone non-increasing. This yields $\lim\limits_{r\rightarrow+\infty}E(r)\in\mathbb{R}$, and so we have $\lim\limits_{r\rightarrow+\infty}E(r)-E(r_{-})=\int_{r_{-}}^{+\infty}E^{\prime}(r)dr=-\int_{r_{-}}^{+\infty}\frac{\beta^2}{r}dr$. The convergence of the latter improper integral leads to
\begin{eqnarray*}
0\leq(1+\varepsilon)^{2}\int_{r_{-}}^{+\infty}\frac{\sin^{2}\theta}{r}dr\leq\int_{r_{-}}^{+\infty}\frac{\beta^2}{r}dr<+\infty.
\end{eqnarray*}

The double inequality $-\frac{3}{2}(r-r_{-})\leq\theta(r)-\theta(r_{-})\leq-\eta(r-r_{-})$, $r\geq r_{-}$, shows that $\lim\limits_{r\rightarrow+\infty}\theta(r)=-\infty$. The function $\theta$ being strictly decreasing, we deduce the existence and uniqueness of the increasing, unbounded from above sequences $(r_{n}^{+})_{n\geq n_{0}}$, $(r_{n}^{-})_{n\geq n_{0}}$, with $r_{n}^{+}>r_{n}^{-}$, given by the formulas
\begin{eqnarray*}
\theta(r_{n}^{+})=\frac{\pi}{4}-2n\pi+\theta(r_{-}),\quad\theta(r_{n}^{-})=\frac{3\pi}{4}-2n\pi+\theta(r_{-}),\quad n\geq n_{0},
\end{eqnarray*}
where $n_{0}\geq1+\left\lceil\frac{\vert\theta(r_{-})\vert}{2\pi}\right\rceil$. Remark also that
\begin{eqnarray*}
\theta(r_{n-1}^{+})=\frac{\pi}{4}-2(n-1)\pi+\theta(r_{-})=\theta(r_{n}^{-})+\frac{3\pi}{2}>\theta(r_{n}^{-}),
\end{eqnarray*}
and, accordingly, $r_{n-1}^{+}<r_{n}^{-}$, $n\geq n_{0}+1$.

We have the estimates
\begin{eqnarray}
\int_{r_{-}}^{+\infty}\frac{\sin^{2}\theta}{r}dr\geq\sum\limits_{n=n_{0}}^{+\infty}\int_{r_{n}^{-}}^{r_{n}^{+}}\frac{1/4}{r}dr\geq\sum\limits_{n=n_{0}}^{+\infty}\frac{r_{n}^{+}-r_{n}^{-}}{4r_{n}^{+}}.\label{estimare_aux_03}
\end{eqnarray}

Taking into account that
\begin{eqnarray*}
-\left[\frac{3}{2}+\frac{c}{(1+\varepsilon)^{\nu}}\right](r_{n}^{+}-r_{n}^{-})&\leq&\theta^{\prime}(\xi_{n})(r_{n}^{+}-r_{n}^{-})\\
&=&\theta(r_{n}^{+})-\theta(r_{n}^{-})=-\frac{\pi}{2}\\
&\leq&-\eta(r_{n}^{+}-r_{n}^{-}),\quad\xi_{n}\in(r_{n}^{-},r_{n}^{+}),
\end{eqnarray*}
which yield
\begin{eqnarray}
\frac{\pi}{2\eta}\geq r_{n}^{+}-r_{n}^{-}\geq\frac{\pi}{3-\frac{2c}{(1+\varepsilon)^{\nu}}},\quad n\geq n_{0},\label{estimare_aux_04}
\end{eqnarray}
we deduce that
\begin{eqnarray*}
r_{m}^{+}&=&r_{n_0}^{-}+\sum\limits_{n=n_{0}+1}^{m}(r_{n}^{-}-r_{n-1}^{+})+\sum\limits_{n=n_{0}+1}^{m}(r_{n}^{+}-r_{n}^{-})\\
&\geq&r_{n_0}^{-}+\sum\limits_{n=n_{0}+1}^{m}(r_{n}^{+}-r_{n}^{-})\geq r_{n_0}^{-}+\frac{\pi}{3-\frac{2c}{(1+\varepsilon)^{\nu}}}(m-n_{0})
\end{eqnarray*}
for any $m\geq n_{0}+1$. Notice the by-product inequalities $2\eta<2<\frac{3+c}{1+c}=3-\frac{2c}{1+c}<3-\frac{2c}{(1+\varepsilon)^{\nu}}$. Further, we have
\begin{eqnarray*}
\frac{\pi}{4}-2n\pi&=&\theta(r_{n}^{+})-\theta(r_{-})=\theta^{\prime}(\mu_{n})(r_{n}^{+}-r_{-})\\
&\leq&-\eta(r_{n}^{+}-r_{-}),\quad\mu_{n}\in(r_{-},r_{n}^{+}),
\end{eqnarray*}
which yield
\begin{eqnarray}
\frac{2n\pi-\frac{\pi}{4}}{\eta}+r_{-}\geq r_{n}^{+},\quad n\geq n_{0}.\label{estimare_aux_05}
\end{eqnarray}

In conclusion, via (\ref{estimare_aux_03})--(\ref{estimare_aux_05}), we obtain that
\begin{eqnarray*}
\int_{r_{-}}^{+\infty}\frac{\sin^{2}\theta}{r}dr&\geq&\sum\limits_{n=n_{0}}^{+\infty}\frac{r_{n}^{+}-r_{n}^{-}}{4r_{n}^{+}}\geq\sum\limits_{n=n_{0}}^{+\infty}\frac{\frac{\pi}{3-\frac{2c}{(1+\varepsilon)^{\nu}}}}{4\left(\frac{2n\pi-\frac{\pi}{4}}{\eta}+r_{-}\right)}\\
&=&\frac{\pi\eta}{12-\frac{8c}{(1+\varepsilon)^{\nu}}}\cdot\sum\limits_{n=n_{0}-1}^{+\infty}\frac{1}{2\pi n+\left(\frac{7\pi}{4}+\eta r_{-}\right)}=+\infty.
\end{eqnarray*}

We have reached a contradiction.

{\bf Polar coordinates III. Escaping $E=0$.\/} The situation depicted in Figure 3 below is inspired by the case of (\ref{vorticitatea_constantin}). There, see \cite{constantin1}, in quadrant I, that is when $\theta\in\left(0,\frac{\pi}{2}\right)+2\pi\cdot\mathbb{Z}$, the algebraic curve $E(\psi,\beta)=\beta^{2}-\frac{4}{3}\vert\psi\vert^{3/2}+\psi^{2}=0$ is concave, the point $(0,0)$ is singular and at the {\lq\lq}smooth peek{\rq\rq} $(\psi_{+},0)$, where $\psi_{+}=\frac{16}{9}$, the curvature is $\frac{9}{4}$. The energy of the non-null equilibria is also negative, $\int_{0}^{u_{i}}f(u)du<0$, where $i\in\{0,1\}$. A consequence of the concavity reads as follows: {\it in quadrant I, each oblique straight line splits the region $E\leq0$ into two disjoint parts, the {\lq\lq}$+${\rq\rq} and the {\lq\lq}$-${\rq\rq}, see Figures 1, 3.\/}

\begin{figure*}[h]
    \centering
        \includegraphics[width=10cm,height=7cm]{C:/octavian/poza3_viena.bmp}\\
        {\pal Figure 3}
    \label{fig:trei}
\end{figure*}

A solution $(\psi,\beta)$ of (\ref{main_ODE}) starting from $(a,0)$ for some very great $a>1$ will reach the ring $1+\varepsilon<R<1+\delta$ in finite time $r$. Since $\psi_{+}>1\geq u_{0}$, the positive quantities $\varepsilon$, $\delta$ can be taken small enough, meaning such that $\psi_{+}>1+\delta>1+\varepsilon>(1+c)^{1/\nu}\geq1$, for the ring to intersect the region $E<0$ of the phase plane as presented in Figure 3. The decay of $E(r)$, detailed previously, has the following consequences: {\it if the point $(\psi(T),\beta(T))$, for some great $T>1$, is on the algebraic curve $E=0$ then the trajectory will cross (transversally) from outside to inside the algebraic curve; if the trajectory will intersect the algebraic curve in its (smooth) peek $(\psi_{+},0)$ then, again, the trajectory will enter the right {\lq\lq}lobe{\rq\rq} of region $E<0$.\/}

Suppose, for the sake of contradiction, that the solution $(\psi,\beta)$ remains {\it for ever\/} outside the region $E\leq0$, that is $\liminf\limits_{r\rightarrow+\infty}E(r)>0$. As in Figure 3, this means its trajectory will intersect the horizontal axis outside the interval $[\psi_{-},\psi_{+}]$. The estimates (\ref{estimare_aux_polar_00}), (\ref{estimare_aux_polar_01}), namely
\begin{eqnarray*}
-\frac{3}{2}-\frac{c}{(1+\varepsilon)^{\nu}}\leq\theta^{\prime}(r)\leq-(1-\lambda)+\frac{1}{2r},\quad r\in[1,+\infty),
\end{eqnarray*}
show that the trajectory will rotate around the origin $O$ infinitely many times and it will also intersect the ring infinitely many times. Taking into account the possible positions of the points from the trajectory that lie inside the ring, we have four (in symmetrical positions) {\lq\lq}siblings{\rq\rq} of the situation depicted in quadrant I. There exist, accordingly, two increasing, unbounded from above sequences $(r_{n}^{+})_{n\geq n_{0}}$, $(r_{n}^{-})_{n\geq n_{0}}$, with $r_{n}^{+}>r_{n}^{-}$, given by the formulas
\begin{eqnarray*}
\theta(r_{n}^{+})=\theta_{1}-2n\pi+\theta(r_{-}),\quad\theta(r_{n}^{-})=\theta_{0}-2n\pi+\theta(r_{-}),\quad n\geq n_{0},
\end{eqnarray*}
for some great integer $n_{0}$. As before, since $2\pi-(\theta_{0}-\theta_{1})>0$, we have $r_{n-1}^{+}<r_{n}^{-}$ when $n\geq n_{0}+1$. Notice also that $R(r)\geq1+\varepsilon>1$ whenever $r\in[r_{n}^{-},r_{n}^{+}]$, see \cite[Eq. (3.18)]{constantin1}. 

When the solution escapes from quadrant I while remaining outside the region $E\leq0$, its possible trajectories in quadrant IV read as in Figure 4. Repeating the argumentation from Polar coordinates II, we reach a contradiction. So, $\lim\limits_{r\rightarrow+\infty}E(r)\leq0$, which means that {\it the solution $(\psi,\beta)$ will enter the region $E\leq0$ in {\lq\lq}finite time{\rq\rq}.\/}

\begin{figure*}[h]
    \centering
        \includegraphics[width=9cm,height=7cm]{C:/octavian/poza4_viena.bmp}\\
        {\pal Figure 4}
    \label{fig:patru}
\end{figure*}

Let us return to the estimates (\ref{estimare_aux_polar_00}), (\ref{estimare_aux_polar_01}). Introduce $\zeta\in(0,1-\lambda)$ such that $r_{++}>\frac{1}{2[(1-\lambda)-\zeta]}>1$. If a solution $(\psi,\beta)$ is still outside the region $E\leq0$ for large values of $r$, meaning for $r\geq\frac{1}{2[(1-\lambda)-\zeta]}$, then we have the estimates $-\frac{3}{2}-\frac{c}{(1+\varepsilon)^{\nu}}\leq\theta^{\prime}\leq-\zeta$ throughout $\left[\frac{1}{2[(1-\lambda)-\zeta]},r_{++}\right)$. The {\lq\lq}$+/-${\rq\rq} splitting of the region $E\leq0$ displayed in Figure 3 implies that, if the solution $(\psi,\beta)$ will not leave quadrant I ever again, then its trajectory will cross from outside to inside the algebraic curve $E=0$ but {\it it will not encounter the origin $O$\/} as the angle $\theta$ is decreasing with non-null {\lq\lq}angular velocity{\rq\rq} $\theta^{\prime}$ ! In conclusion, if the solution $(\psi,\beta)$ has not reached the origin $O$ in a short period of the time $r$ then the only possible {\lq\lq}entry windows{\rq\rq} are quadrants II and IV. See Figure 4.

{\bf Transversality of flow on the vertical axis.\/} Assume again that the energy of the non-null equilibria is negative, $\int_{0}^{u_{i}}f(u)du<0$, where $i\in\{0,1\}$, and the algebraic curve $E=0$ intersects the vertical axis $[\psi=0]$ only in the singular point $O$. If the solution $(\psi,\beta)$ of (\ref{main_ODE}) is outside the region $E\leq0$ then we have
\begin{eqnarray*}
\psi^{\prime}=\beta\neq0,\quad\beta^{\prime}=-\frac{\beta}{r},\quad r\geq1,
\end{eqnarray*}
whenever $\psi(r)=0$. Take $r_{\star}>1$ such a zero of the function $\psi$. Applying the local inversion theorem, we deduce the existence of a  $C^1$--function $r=r(\psi)$ with $r:(-\varepsilon_{\star},\varepsilon_{\star})\rightarrow(r_{\star}-\delta_{\star},r_{\star}+\delta_{\star})\subset(1,+\infty)$ for some small $\varepsilon_{\star}$, $\delta_{\star}>0$. We have now a new ODE for characterizing the other function $\beta=\beta(r(\psi))=\beta(\psi)$, namely
\begin{eqnarray*}
\left\{
\begin{array}{ll}
\frac{d\beta}{d\psi}=\frac{\beta^{\prime}}{\psi^{\prime}}=-\frac{1}{r(\psi)}-f(\psi)\cdot\beta^{-1},\quad\psi\in(-\varepsilon_{\star},\varepsilon_{\star}),\\
\beta(0)=\beta_{\star}\neq0.
\end{array}
\right.
\end{eqnarray*}

The function $\beta\mapsto\frac{1}{\beta}$ being locally Lipschitzian in $\mathbb{R}-\{0\}$, the solution of this new ODE exhibits continuous dependence of the data. The same issue has been dealt with in \cite[Eq. (3.12)]{constantin1} by means of polar coordinates. The topological arguments from \cite[Section 3]{constantin1} can now be applied verbatim to establish that {\it there exists some $a>1$ great enough in order to have a solution $(\psi,\beta)$ that will reach the null equilibrium $O$ in finite time $r$.\/}

\section{An example of vorticity function}
Introduce the function
\begin{eqnarray*}
f(u)=u-g(u)=\left\{
\begin{array}{ll}
u-\frac{u}{\sqrt{\vert u\vert}}\left[1+c_{1}-\sin\left(\frac{c_{2}u^{2}}{u^{2}+1}\right)\right],\quad u\neq0,\\
0,\quad u=0,
\end{array}
\right.
\end{eqnarray*}
where $0<c_{1}=\sin\frac{c_2}{2}<c_{2}<\frac{3-2\sqrt{2}}{4+3\sqrt{2}}<0.02$.

First of all, notice that for any $\varepsilon$ in a small interval centered in zero, the algebraic curve $E^{\varepsilon}=\beta^{2}+\psi^{2}-(1+\varepsilon)\cdot\frac{4}{3}\vert\psi\vert^{\frac{3}{2}}=0$ is concave in quadrant I and with the smooth peek $\psi_{+}^{\varepsilon}=\frac{16}{9}(1+\varepsilon)^{2}$. Notice further that the energy associated to the flow of $f$ can be estimated by
\begin{eqnarray*}
E(\psi,\beta)&=&\frac{\beta^2}{2}+\int_{0}^{\psi}f(u)du\\
&\leq&\frac{\beta^{2}+\psi^{2}}{2}-[1-(c_{2}-c_{1})]\cdot\frac{2}{3}\vert\psi\vert^{3/2}
\end{eqnarray*}
and respectively
\begin{eqnarray*}
E(\psi,\beta)\geq\frac{\beta^{2}+\psi^{2}}{2}-(1+c_{1})\cdot\frac{2}{3}\vert\psi\vert^{3/2}.
\end{eqnarray*}
So, lying in between the curves $E_{c_1}=0$ and $E_{c_{1}-c_{2}}=0$, the algebraic curve $E=0$ will obey the {\lq\lq}$+/-${\rq\rq} splitting condition.

Second, let us find the zeros of $f$. Obviously, they are fixed points of $g$. Besides the null solution, the algebraic equation $g(u)=u$ reads as
\begin{eqnarray*}
1+c_{1}-\sin\frac{c_{2}u^{2}}{u^{2}+1}=\sqrt{\vert u\vert},\quad u\neq0.
\end{eqnarray*}
In $(0,+\infty)$, the left-hand member of the equation is a decreasing function while the right-hand member is increasing (strictly). In consequence, the equation can have at most one solution here. Since $c_{1}=\sin\frac{c_2}{2}$, the solution reads as $u_{0}=1$. Similarly, we find the third fixed point $u_{1}=-u_{0}$.

Third, to find the quantity $\eta(a)$ that describes the growth of the vorticity function $f$ around $a$, notice that
\begin{eqnarray*}
\vert f(u)\vert\leq\vert u\vert+\sqrt{\vert u\vert}(1+c_{1}+c_{2})\leq\vert u\vert+\sqrt{\vert u\vert}\left(1+\frac{3c_{2}}{2}\right)
\end{eqnarray*}
and so the inequality $\vert f(u)\vert\leq\eta a$ will reduce to $\left(1+\frac{3c_{2}}{2}\right)^{2}\leq\frac{\left(\frac{3\eta}{4}-1\right)^{2}}{1+\frac{\eta}{4}}$. The function $\eta\mapsto h(\eta)=\frac{\left(\frac{3\eta}{4}-1\right)^{2}}{1+\frac{\eta}{4}}$ is increasing in $\left(3,\frac{7}{2}\right]$. By noticing the elementary inequality $\frac{2}{3}\left(\sqrt{\frac{27}{22}}-1\right)>\frac{3-2\sqrt{2}}{4+3\sqrt{2}}$, we deduce that $1+\frac{3c_{2}}{2}<\sqrt{\frac{27}{22}}-1<\sqrt{h\left(\frac{10}{3}\right)}$, meaning that $\eta(a)=\frac{10}{3}$.

Fourth, to find the coefficient $L(a)$ of the locally Lipschitzian (outside $\{0\}$) function $g$, we have the estimates
\begin{eqnarray*}
\vert f^{\prime}(u)\vert&\leq&1+\frac{1}{2\sqrt{\vert u\vert}}\cdot(1+c_{1}+c_{2})+2c_{2}\cdot\frac{\vert u\vert^{\frac{3}{2}}}{(u^{2}+1)^{2}}\\
&=&1+\frac{1+c_{1}+c_{2}}{2\sqrt{\vert u\vert}}+2c_{2}\cdot\frac{\vert u\vert^{\frac{3}{2}}}{(u^{2}+1)^{\frac{3}{2}}}\cdot\frac{1}{\sqrt{u^{2}+1}}\\
&\leq&1+2c_{2}+\frac{1+c_{1}+c_{2}}{2\sqrt{\left(1-\frac{\eta}{4}\right)a}}\leq1+2c_{2}+(1+c_{1}+c_{2})\sqrt{\frac{2}{a}}\\
&<&1+2c_{2}+\left(1+\frac{3c_{2}}{2}\right)\sqrt{2}<\frac{5}{2}.
\end{eqnarray*}
The latter estimate is equivalent to $c_{2}<\frac{3-2\sqrt{2}}{4+3\sqrt{2}}$.

Fifth, given $\psi>0$, observe that
\begin{eqnarray*}
\int_{0}^{\psi}g(u)du&\geq&\int_{0}^{\psi}\sqrt{u}du\cdot(1+c_{1}-\sin c_{2})=\frac{\psi^{3/2}}{\frac{3}{2}\cdot\frac{1}{1+c_{1}-\sin c_{2}}}\\
&=&\frac{\psi g(\psi)}{2\lambda},
\end{eqnarray*}
where $\lambda=\frac{3}{4}\cdot\frac{1}{1+\sin\frac{c_2}{2}-\sin c_{2}}\in\left(\frac{3}{4},1\right)$.

Sixth, regarding the quantities $c$ and $\nu$ from (\ref{estimare_aux_polar_01}), observe that
\begin{eqnarray*}
0&\leq&\frac{\vert\psi\vert^{3/2}}{R^2}[(1-\sin c_{2})+c_{1}]\\
&\leq&\frac{\psi g(\psi)}{R^2}=\frac{\vert\psi\vert^{3/2}}{R^2}\left[\left(1-\sin\frac{c_{2}\psi^{2}}{\psi^{2}+1}\right)+c_{1}\right]\\
&\leq&\frac{\vert\psi\vert^{3/2}}{R^2}(1+c_{1})\leq\frac{\vert\psi\vert^{3/2}}{R^2}\left(1+\frac{c_{2}}{2}\right),
\end{eqnarray*}
and so $c=\frac{c_2}{2}$, $\nu=\frac{1}{2}$.

\end{document}